\documentclass[letterpaper]{article}
\usepackage{jheppub_mod}

\newcommand{\beq}{\begin{eqnarray}}
\newcommand{\eeq}{\end{eqnarray}}

\begin{document}

\begin{center}
{\huge \textbf{Vector-like quarks and heavy coloured bosons at the LHC}}

\vskip 1cm
\end{center}

\vskip0.2cm

\begin{center}
\textbf{{A.Deandrea$^{1,2}$, A.M. Iyer$^{3,4}$} \vskip 8pt }

{\small
$^1$\textit{Universit\'e de Lyon, France; Universit\'e Lyon 1,
CNRS/IN2P3, UMR5822 IPNL,\\ F-69622 Villeurbanne Cedex, France}\\[0pt]
\vspace*{0.1cm} $^2$\textit{Institut Universitaire de France, 103 boulevard
Saint-Michel, 75005 Paris, France}\\[0pt]
\vspace*{0.1cm} $^3$\textit{INFN-Sezione di Napoli, Via Cintia, 80126 Napoli, Italia}\\[0pt]
\vspace*{0.1cm} $^4$\textit{Department of Theoretical Physics,
Tata Institute of Fundamental Research,
Mumbai, India}\\[0pt]
}
\end{center}

\vskip 1cm
\begin{abstract} 
\noindent
We investigate the production of heavy coloured scalars and vectors  and their relevance at LHC for the study of vector-like quarks ($T$). These coloured states ($C$) 
are present in a large number of extensions of the standard model, in particular in composite models and in extra dimensional models. Assuming that 
these bosonic states are heavier than the vector-like quarks (VLQ), we consider their production through the process $p~p\rightarrow C\rightarrow tT$. Large QCD production cross-sections for $C$ enable us to probe heavier masses for the VLQ and thereby allowing to put stronger limits on the 
vector-like quarks which are produced in their decay chain. We adopt a universal analysis strategy by including leptons under the classification of `jets', 
thereby limiting the bias towards a specific combination of final state. We also study the possibility of disentangling these scenarios from supersymmetric 
extensions of the Standard Model by using simple discriminants based on jet multiplicity and missing energy. We demonstrate that a simple set of cuts are 
sufficient to disentangle the VLQ signal from the backgrounds. In models with a moderate $B.R.(C\rightarrow Tt)$, the analysis enables one to get a hint of VLQ masses as heavy as 3 TeV.
\end{abstract}

\newpage
\tableofcontents

\section{Introduction}
The Standard Model (SM) and its mechanism for the electroweak symmetry breaking is a quite successful description of the observed data. 
However a more fundamental description of the electroweak and strong interactions is still missing and many possible avenues are open in this respect.
Two popular possibilities for physics beyond the SM (BSM) suggest to consider the Higgs boson either as a fundamental scalar particle (as for example in 
supersymmetric  extensions of the SM) or as a composite particle (as in composite models). This second possibility is usually described at the effective 
Lagrangian level, but if a more fundamental theory in terms of constituent fermions is considered, the presence of other bound states together with the 
Vector-Like Quarks (VLQ) is typically unavoidable. The common assumption in effective composite Higgs models is that all these extra states are much heavier. 
This can be indeed the case, however simple models implementing the composite idea at the level of constituent new fermions can also allow composite 
states of similar mass range as the VLQs.  These new states can be even lighter if protected by a symmetry.
Typically independent studies are conducted for investigating the discovery potential of either the coloured bosons or the vector-like quarks. Possibility of 
composite coloured scalars have been discussed in various extensions of the SM like \cite{Lewis:2011zb,Cacciapaglia:2015eqa}. The coloured (composite) 
states can also manifest as spin-1 states generally corresponding to the KK-excitations of gluons in warped extra-dimensional models 
\cite{Randall:1999ee,Gherghetta:2000qt}. Collider studies for these coloured vectors have been considered in 
\cite{Agashe:2006hk,Lillie:2007yh,Guchait:2007jd,Iyer:2016yjb}. In parallel, comprehensive studies studies for the search of the vector-like top partners, 
in both the composite and RS-like scenarios have been performed in 
\cite{Han:2003wu,Contino:2006qr,Carena:2006jx,Matsumoto:2008fq,Anastasiou:2009rv,Kribs:2010ii,Gopalakrishna:2011ef,DeSimone:2012fs,
Vignaroli:2012sf,Gopalakrishna:2013hua,Buchkremer:2013bha, Banfi:2013yoa, Li:2013xba, Gripaios:2014pqa,Chala:2014mma, Endo:2014bsa, Dolan:2016eki}. 
 
Recently, there have been efforts to consider a unified search strategy for the heavy bosons and the VLQ. Typically it is assumed that the bosons are heavier 
than the VLQ, such that they decay into a VLQ and SM quark (often third generation is assumed). In particular VLQs stemming from heavier states and decaying 
to multi-top final states were considered in \cite{Deandrea:2014raa}. The first search for a spin-1 decaying into a $Tt$ final state was performed by ATLAS 
\cite{Sirunyan:2017bfa} at 13 TeV with 2.6 $fb^{-1}$ of data. Phenomenological analysis for similar topologies were considered for a heavy 
scalar $H$ decaying into a $Tt$, suggesting kinematic variables to uniquely identify the presence of a VLQ state \cite{Iyer:2016qzd}.

The phase space for the search of VLQ through such channels can be considerably extended if the VLQ is produced in the decay of a heavier state. The 
non-coloured states suffer a substantial reduction in the production cross-sections at high mass. The corresponding effect on the coloured excitations 
is comparatively much less and presents an avenue to probe the deeper realms of the VLQ masses which would be difficult to explore from direct search 
channels. A strategy to this effect was suggested in \cite{Chala:2014mma} with a heavy gluon and a VLQ, discussing different simplified models and search 
strategies. In the following we shall focus on the possibility of new coloured states heavier than the VLQs, in order to show how the LHC reach in the searches 
of VLQs can be extended. Depending on the model framework, the identity of the final state will be different. We present a unified search strategy by considering 
jets and not subjecting the leptons to the standard isolation criteria. As a result even the leptons (and photons) are pushed into the classification of jets and this 
analysis is applicable across all kinds of models with a similar particle content. This in turn gives 
bounds which can be applied directly on the composite Higgs models which are considered in the literature.

The paper is organised as follows: in section 2 we briefly discuss the effective Lagrangian formalism and the different cases which will be discussed. In section 3 
we perform a simplified analysis at the LHC in order to show the potential of these searches and discuss the possibility to discriminate such a signal from 
background and other cases giving similar final states. In section 4 we give our conclusions. A short appendix contains the details of the numerical simulation.

\section{Effective Lagrangian description}
Fundamental composite models typically contain a large number of resonances: scalars, fermions and spin-1 particles.
We  discuss the effective Lagrangian description following two scenarios characterised by the nature of the particle content.\\

Case A: Heavy scalar and the fermionic VLQs.\\

To begin with, we consider the heavy scalar ``mesons" and the fermionic ``baryons" VLQs \footnote{In the following we shall refer improperly to ``mesons" 
and ``baryons" as respectively two and three fermion bound states. The correct labelling of the composite states depends on the representation and on the 
type of bound state they form, as the force bounding the fundamental fermions is not necessarily QCD-like. Finally the obvious and correct criterium is if they 
carry a baryon number or not. However we shall not use this classification as it requires to specify the details of the particular model, while we aim 
at a generic description.}. 
The coloured pseudo Nambu-Goldstone Bosons are expected to have a mass which can be in the same 
range as the top partners, therefore their phenomenology crucially depends on the mass hierarchy. We assume here, that the VLQs are lighter, therefore 
allowing their production in the decay chain of the coloured bosons.
As an example in the model considered in \cite{Cacciapaglia:2015eqa}, the spectrum contains a complex colour sextet (with charge $Q=4/3$) 
and a real colour octet. The presence of such states is a generic prediction of any dynamics that also generates the VLQs.

The couplings of the sextet and octet can be written considering the invariance under colour and charge at the effective Lagrangian level. 
The octet can only couple to a quark-antiquark pair. A sextet on the contrary can only be obtained by combining two quarks, as for the SU(3) representations 
${\bf 3} \otimes {\bf 3} \supset {\bf 6}$, or four anti-quarks ${\bf \bar{3}} \otimes {\bf \bar{3}} \otimes {\bf \bar{3}} \otimes {\bf \bar{3}} \supset {\bf 6}$.

The effective Lagrangian is given as \cite{Cacciapaglia:2015eqa}
\beq
\mathcal{L}_{eff} & = &{|D_\mu\pi_6 |}^2 - m_{\pi_6}^2 {|\pi_6|}^2 + \frac{1}{2} {(D_\mu \pi_8)}^2 - \frac{1}{2} m_{\pi_8}^2 {(\pi_8)}^2 - V_{\rm scalar} (\pi_6, \pi_8) \nonumber \\
& & + b_R ~ \pi_6 t_{R}^c t_{R}^c + b_L ~ \pi_6^c t_{L} t_{L} + c~ \pi_8 t_R^c t_L + h.c. \label{eq: lag sextet}
\eeq
where $t_{L/R}$ are chiral Weyl spinors (${^c}$ indicates the charge conjugation) and $V_{\rm scalar}$ is a generic self-interactions between the scalars.
Parity is not conserved in \ref{eq: lag sextet}, because only the coupling $b_R$ corresponds to a gauge invariant operators. The other couplings are generated 
only at the electro-weak symmetry breaking. 

Note that we couple the sextet only to the third generation quarks, as is typically assumed in composite models based on partial compositeness considerations, 
but in principle a coupling to the other quark generations is also possible.\\

Case B: Heavy Vector and the fermionic  VLQs: \\

This is typically characteristic of spectrum in warped extra-dimension models. For simplicity we consider the VLQ to be a singlet under SU(2). However, as will 
be described later, the analysis will proceed independently of any specific final state and is fairly robust. The effective Lagrangian for particle content under 
consideration is given as:
\begin{equation}
\mathcal{L}_{RS}\supset \bar qT^a\gamma^\mu qG^a_\mu+\left( Y_t\bar Q_3H  t_p+M_{t'}\bar t_pt_p + Y_t\bar Q_3H t+h.c\right)
\label{effective_lagrangian}
\end{equation}
Here $G_\mu$ represents the first KK excitation of the gluon in RS-like models and $H$ is the SM Higgs doublet. For the specific case under consideration, 
the vector like state $t_p$ mixes with the top quark leading to two eigenstates $(t_1,\, t_2)$, where $t_1$ is identified as the SM top and $t_2$ is the top partner.
Since it is a singlet SU(2), $t_2$ has only three decay modes: $Wb,Zh,th$ with the $Wb$ constituting $50\%$ of the total branching fraction.

\subsection{Spin-1, Sextet and Octet masses}
The mass terms for the vectors, coloured sextet and octet mesons were described in detail in \cite{Cacciapaglia:2015eqa} in the case of the global symmetries
$SU(4)$ and $SU(6)$ for the fundamental fermions giving rise to the VLQ ``baryons'', the various pNGB ``pions'' and the spin-1 vectors. The spin-1 states 
are allowed to have an $SU(6)$ invariant mass term which is typically a heavy mass scale. The $SU(6)$ symmetry is also broken explicitly by a mass term, 
which gives the mass ballpark for the coloured pNGBs. For the VLQs an extra mass term is also present, therefore the naive expectation is that the mass of 
the VLQs will receive both a contribution from the dynamical and explicit $SU(6)$ breaking, while the coloured pNGBs will only receive a contribution from 
the explicit breaking. The composite spin one resonances, due to the $SU(6)$ invariant mass term are typically expected to be heavy.

This picture can be modified when the model is close to the conformal window, where large anomalous dimensions may be generated for  
some of the composite operators. In the following we shall consider only the effective model for the bound states and treat the mass terms as parameters 
of the model, without referring to a particular fundamental dynamics behind them (which is anyway model dependent and which requires detailed lattice 
simulations in order to establish the spectrum of the low energy theory).
The situation we examine is the one in which the coloured sextet and octet mesons as well as the spin-1 resonances are heavier than 
the VLQs top partners, allowing the decay modes of the $\pi_6$, $\pi_8$ and spin-1 resonances to the VLQs top partners. 

\subsection{Pair vs single production of heavy coloured bosons and LHC bounds}

The possibility of new massive colour octet vector bosons with sizeable decay into the VLQs was considered by \cite{Araque:2015cna} and the impact on the 
LHC searches was assessed using as an example a simplified version of the composite Higgs model, based on the $SO(5)/SO(4)$ coset and a composite 
right-handed top quark \cite{Agashe:2004rs,DeSimone:2012fs}. The focus in \cite{Araque:2015cna} was on pair production of the VLQs. The present limits 
on VLQ masses are however higher and single production of the VLQs, even if more model dependent, becomes more constraining as the single production 
cross-section becomes larger than pair production for large enough masses of the VLQs. In the following we shall therefore focus on single production of the 
VLQs in the decay chain of coloured vector bosons, sextet and octet coloured mesons.

The present limits on VLQs come from the detailed searches of their decay products (both for pairs and single production) assuming either a 100\%
branching in one specific decay mode or a set of decay modes (typically those into a SM quark plus the Higgs boson or the W, Z gauge bosons).
The lower bounds for the masses are typically in the TeV mass range. We shall use them as a guideline, but lower mass values can be
possible if extra decay modes are present or dominant. For example top partners with masses below 900 GeV are excluded independently of 
whether they decay into $Zt$, $Ht$, $Wb$ or $St$ (where $S$ gives either missing energy or $b$-quarks) or any combinations of those \cite{Chala:2017xgc}.

Experimental searches using the VLQ single production from heavy coloured gluon were recently performed in the 4 $b$-quarks final state
\cite{Aad:2016shx}. In that case stringent limits on the production cross section times branching ratio can be set. This kind of analyses can be 
generalised to other final states and to different heavy coloured objects. Indeed, while colour octets are typically considered, colour sextets are not.
In the following we shall consider those heavy states (including the sextet) as a mechanism to produce or bound vector-like top partners $T$ in association 
with a SM top quark.

\section{Collider Analysis at the LHC}
In this section we explain the details of the collider strategy employed in our analysis. Though there are two possibilities for the spin and the representation of the coloured objects, the analysis proceeds in a democratic fashion without any bias on the nature of the coupling. For simplicity they will be collectively referred to as $C$. As explained earlier, large production cross sections 
for the coloured states not only enable us to probe heavier masses for such states, but also serve as a portal to the existence of heavy vector like states. 
We consider the production of the these states (as a typical feature of the composite models we consider a top-partner $T$) in association with a SM top 
through the following process:
\begin{equation}
	p~p\rightarrow C\rightarrow t~T
\end{equation}
The electric charge of $T$ is the same as that of the SM top (top-partners), however, depending on the representation of $T$ under ($SU(3)$, $SU(2)$), the 
final states are different. The following two cases are discussed:
\begin{itemize}
	\item Case A - $(\bar 3, 1)$: They are produced in association with tops of the same sign through coloured objects which transforms as a sextet. 
	The only possible decay mode is $T\rightarrow t g$.
	\item Case B - $(3,1)$: They are produced in association with opposite sign tops though coloured octets. The dominant decay mode is $T\rightarrow W b$ 
	which accounts for 50$\%$ of the branching fraction. The subdominant modes include $Zt$ and $th$ each of which accounts for 25$\%$.
\end{itemize}
Though the final states in the above two scenarios are different, they are characterised by the presence of three four-vectors at the particle level.
Typically in a collider, final states are characterised by combinations of jets, leptons, photons and missing energy. Heavier objects like $W,Z,t$ are 
reconstructed from their final state decay products.
With increasing boost of the particles, decay products of objects like $t,W,Z$   are concentrated in small annular region typically of radius 
$\Delta R\sim 2m/p_T$, where $m$ is the mass of the corresponding object having transverse momentum $p_T$. As a result these objects also fall 
into the broad classification of `jets'  and can then be considered on the same footing as jet initiated by a gluon or a light quark to begin with.
In order to put this into practice, it is essential that the size of the jet-reconstruction radius is such that the decay products of the heavier objects (especially 
the top) are captured and depend on the transverse momentum. As an illustration we consider the following two benchmark points:
\begin{equation}
	\text{BP1:} m_{C}=3~\text{TeV}~m_T=1.7~\text{TeV}\;\;\;\;\;\;\;\;\;\;\;\text{BP2:} m_{C}=3.5~\text{TeV}~m_T=2.5~\text{TeV}
	\label{bp}
\end{equation}
\begin{figure}[!htb]
	\begin{center}
		\begin{tabular}{cc}
			\includegraphics[width=7.2cm]{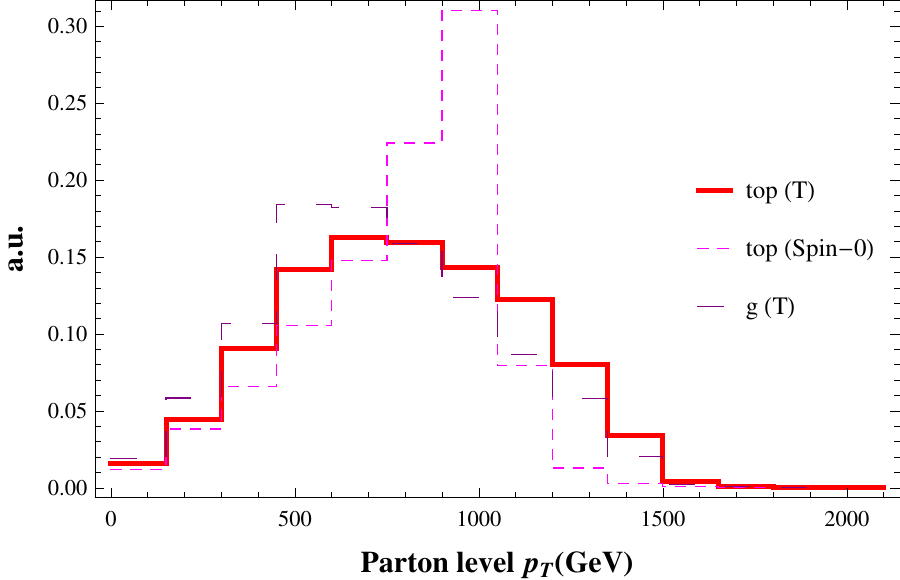}&\includegraphics[width=7.2cm]{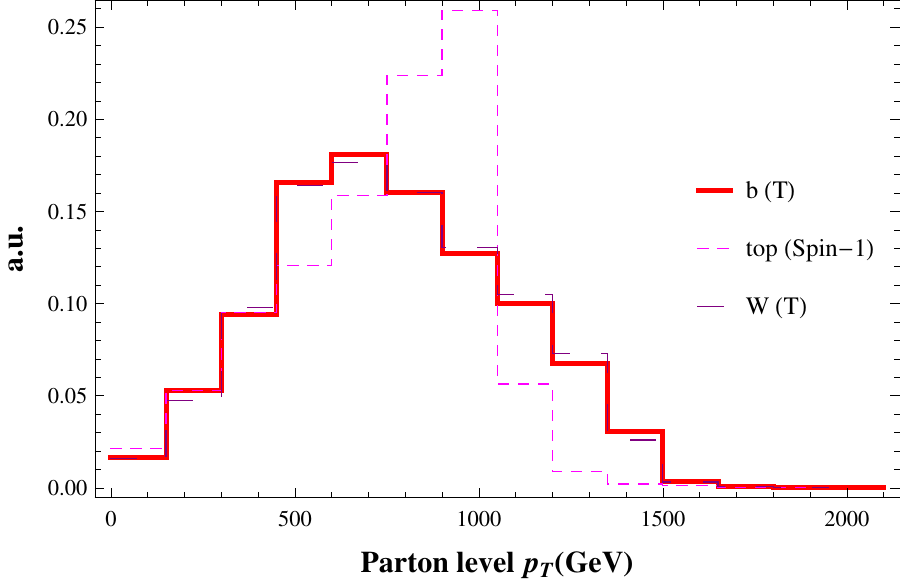}\\
			\includegraphics[width=7.2cm]{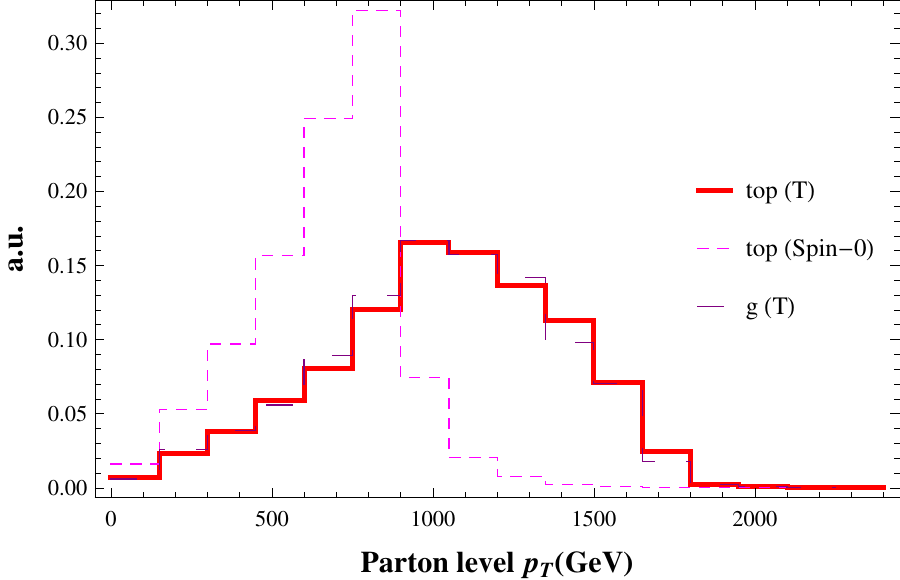}&\includegraphics[width=7.2cm]{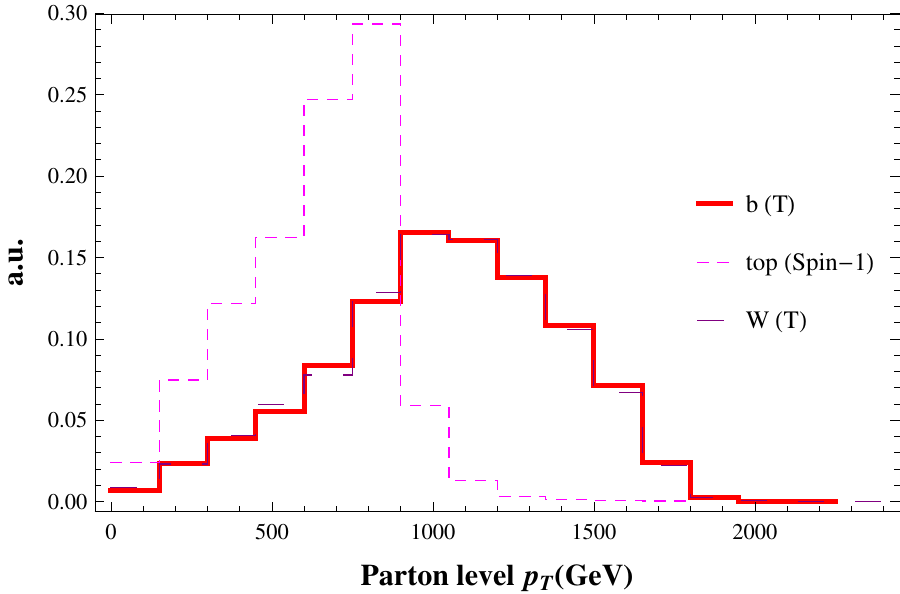}
		\end{tabular}
	\end{center}
	\caption{ The left column gives the parton level $p_T$ for $ttg$ final state while the right column gives the corresponding $p_T$ for $tWb$ final state.
		The top row corresponds to the BP1 while the bottom row corresponds to BP2.}
	\protect\label{parton}
\end{figure}  
Fig. \ref{parton} gives the parton level $p_T$ for the two benchmark points originating from either the C or the T.
The plots in the right column correspond to the $ttg$ final state (CASE A) and left column correspond to the $tWb$ final state (CASE B).
It can be clearly seen that a choice of $R=0.8$ for forming these jets is sufficient to capture all the information from the heavier objects. 
Appendix \ref{app:A} gives the details of the simulation employed in our analysis. 

\subsection{Background Discrimination} 
\begin{figure}[!htb]
	\begin{center}
		\begin{tabular}{cc}
			\includegraphics[width=7.2cm]{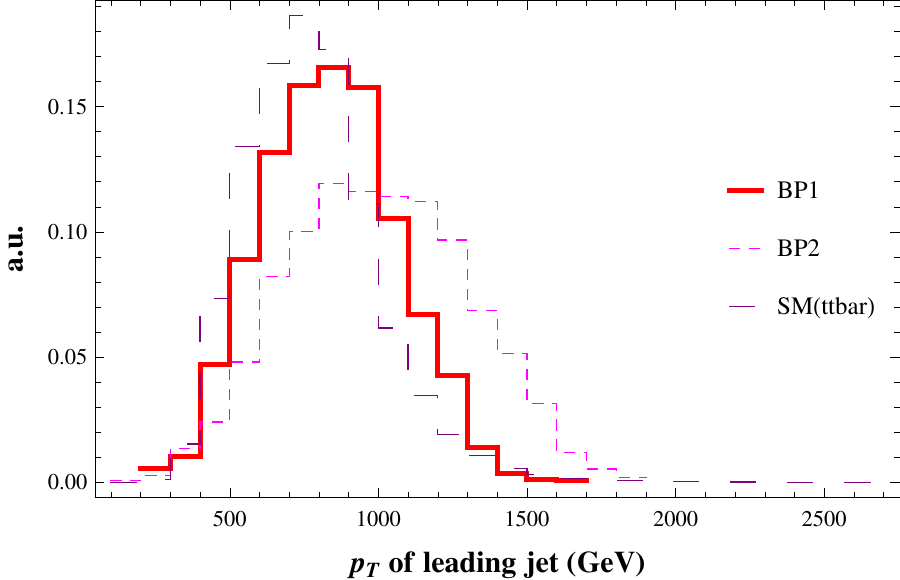}&\includegraphics[width=7.2cm]{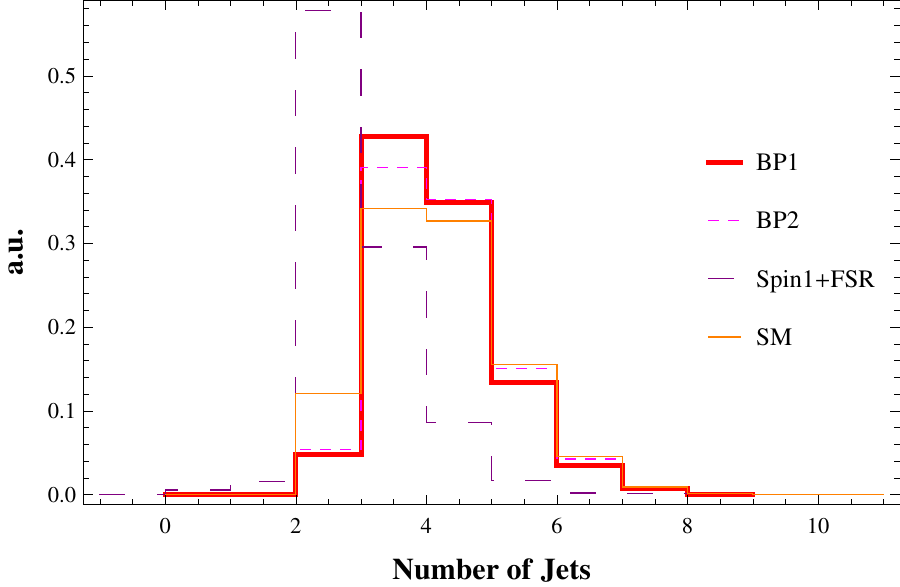}
		\end{tabular}
	\end{center}
	\caption{ $p_T$ of the leading jet for the signal and two bench-mark points.}
	\protect\label{ptjetno}
\end{figure}

In view of identifying a unique signature for the presence of VLQs, it is essential to identify different processes which may impersonate the signal. These 
include both SM and other NP processes and are discussed below:\\
\textbf{$t\bar t + jets$:} In accordance with the event selection criteria explained in \ref{app:A}, $t\bar t + jets$ constitutes the most dominant SM background. The matrix element for the process is given as 
\begin{equation}
	\mathcal{M}_{t \bar t +jets}=\mathcal{M}(pp\rightarrow  t\bar t)+\mathcal{M}(pp\rightarrow t\bar t+j)+\mathcal{M}(pp\rightarrow t\bar t+j+j)
\end{equation}
where $j$ is a parton.
As seen in Fig. \ref{parton}, the signal kinematics will be accompanied with high $p_T$ partons. In order to populate the phase space in the signal regime, 
the events are simulated by requiring the minimum invariant $p_T$ to be 800 GeV. Left panel of Fig. \ref{ptjetno} gives a comparison of $p_T$ of the leading jet for the 
background and the two benchmark points. Though BP1 and BP2 have been plotted for the $ttg$ final state,  $tWb$ or any final state from a similar 
NP masses will also exhibit the same behaviour.
Since all the three plots have a similar pattern, they suggest that the background has been fairly well represented 
in the signal phase space.  The effective cross section is reduced to 1740 fb and we simulate 0.5 million events for the same.\\
\textbf{$C\rightarrow t\bar t$:} In the presence of a heavy coloured object, $t\bar t$ final state constitutes its most favourable decay mode and must be 
accounted for. We generate these events by also considering the emission of a parton from the final state top. The total matrix element for the process 
is given as
\begin{equation}
	\mathcal{M}_{C=tt}=\mathcal{M}(pp\rightarrow C\rightarrow t\bar t)+\mathcal{M}(pp\rightarrow C\rightarrow t\bar t+j)
\end{equation}

\textbf{SUSY processes:} Minimal supersymmetric extensions of the SM are characterised by the absence of these vector like states. However certain decays 
of coloured objects like the stops or the gluinos may potentially impersonate the VLQ signal, either in terms of jet multiplicity or the number of isolated leptons.
In general there are a plethora of processes in MSSM which may exhibit similar kinematics or multiplicity of certain combinations of final state collider objects. 
As a representative, we consider the following two processes:\\
a) Pair production of stops: We consider the pair production of 1 TeV stops with the decay $\tilde t \rightarrow t \chi^0_1$. The final state in this case is 
$tt+ E^{miss}_T$. The visible final states in this case is similar to the SM $t \bar t$ + jets and the $C\rightarrow t\bar t$.\\
b) Pair production of gluinos: Searches for the gluinos in the 1l+jets + $E_T^{miss}$ final state restrict the mass of the gluinos to be $m_{\tilde g}\gtrsim 2$ TeV.

\begin{figure}[!htb]
	\begin{center}
		\begin{tabular}{cc}
			\includegraphics[width=7.2cm]{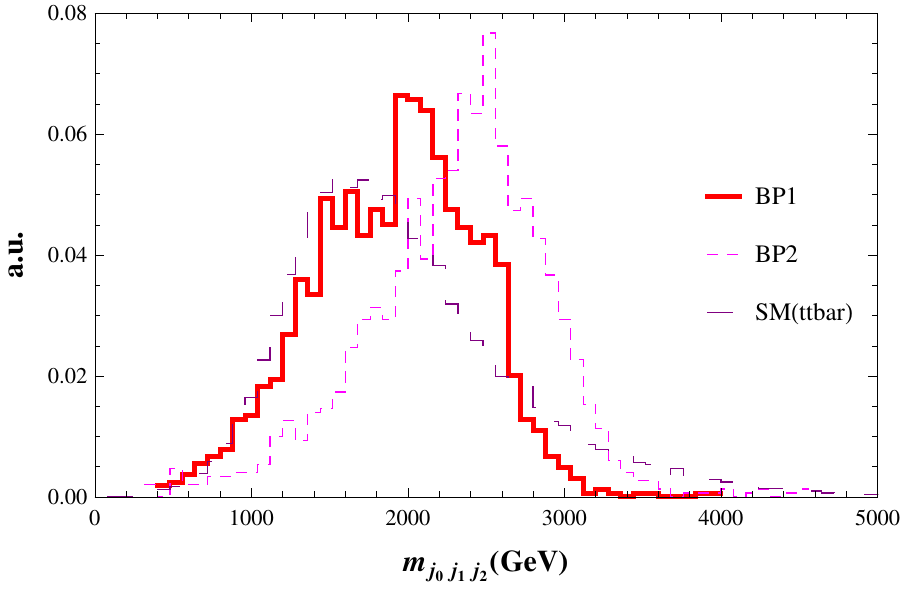}&	\includegraphics[width=7.2cm]{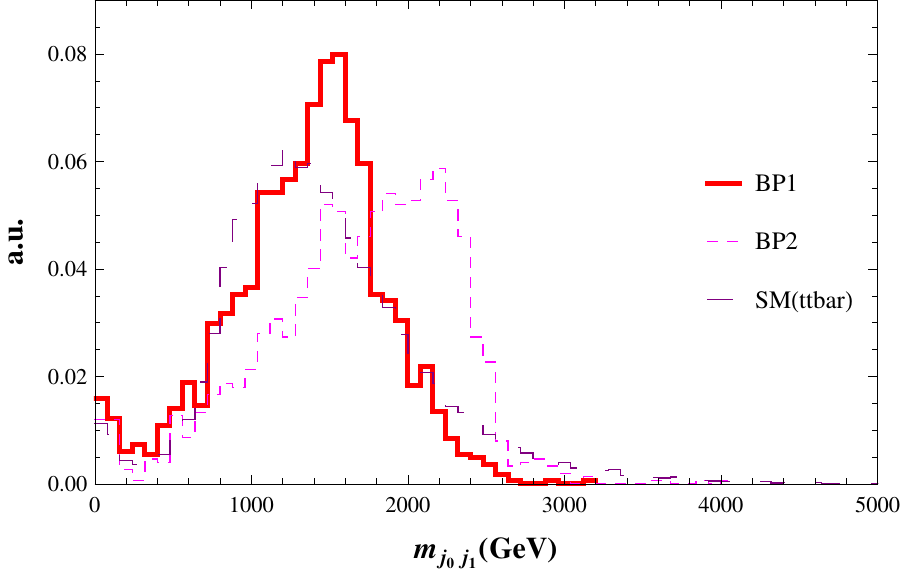}
		\end{tabular}
	\end{center}
	\caption{ The total invariant mass of the three leading jets (left) and the ivariant mass of the two leading jets (right).}
	\protect\label{invmass}
\end{figure}

\begin{figure}[!htb]
	\begin{center}
		\begin{tabular}{c}
			\includegraphics[width=7.2cm]{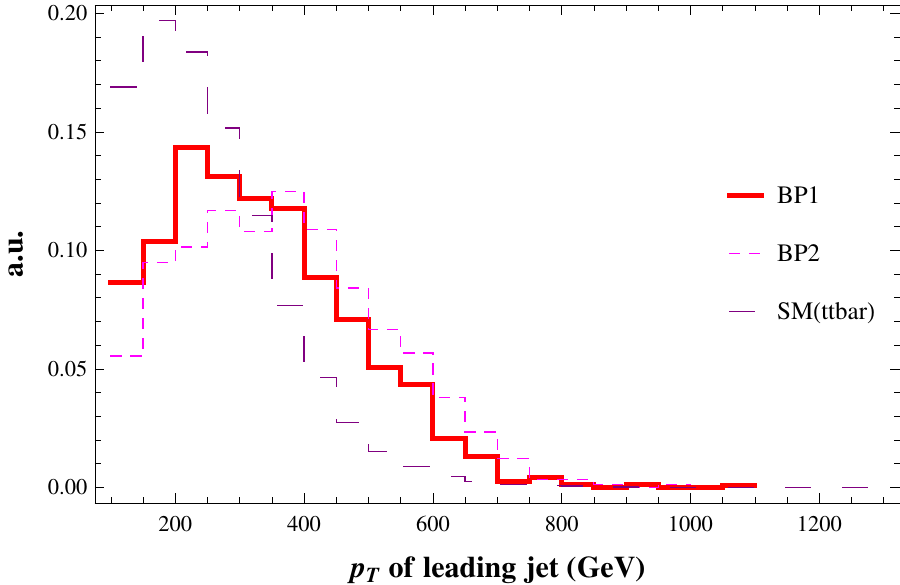}
		\end{tabular}
	\end{center}
	\caption{ $p_T$ distribution for the third leading jet}
	\protect\label{pt3}
\end{figure}

All these processes will now be collectively referred to as  `background' to our spin-1+VLQ decay topology. We now list the different variables which will be 
useful in distinguishing the VLQ signal from these background processes.\\
\begin{itemize}
	\item Jet~multiplicity $(n_J)$: The signal is characterised by a peak at 3 in the jet distribution as shown in right panel of Fig.\ref{ptjetno} . The other NP 
	processes are however characterised by smaller ($C\rightarrow t\bar t$ or $\bar t\bar t$) or larger ($\tilde g\tilde g$) jet multiplicities. While the SM 
	process exhibits a similar nature, it can be substantially reduced by a minimum requirement on the leading jet $p_T$. 
	
	\item $p_T$ of the leading jet ($p^{(0)}_T$): For both the benchmark points in Eq. \ref{bp}, the most energetic partons originate from the spin-1 vertex and are likely to constitute the leading jet. Fig \ref{ptjetno} left panel gives the $p_T$ distribution for the background process (left) and the signal (right). One can expect the $C\rightarrow t\bar t$ to have a similar distribution. However, this is useful in eliminating the SM and the SUSY background.
\item Reconstruction of the $W$ boson mass: For the topology under consideration we demand that the missing momenta is only due to the neutrino. The $z$ component of the neutrino is explicitly evaluated by assuming that it to be originating from the $W$. This is particularly useful in keeping SUSY backgrounds at bay.
\item Total invariant mass $m_{j_0j_1j_2}$: Owing to the presence of missing energy from the neutrino, the total invariant mass of the three leading jets is likely to exhibit an edge at the mass of the spin-1 resonance. Left plot of Fig. \ref{invmass} gives the distribution for this variable for the SM backgroud and the two benchmark ppints.
 It is clear that as the mass of the colured resonance increases the, discriminatory capability of this variable becomes even more stronger.
	\item $p_T$ of the third leading jet ($p^{(2)}_T$): The partons from $T$ are also considerably energetic and will result in the corresponding jets with large $p_T$ and  generally constitute the sub-leding jets. Fig.\ref{pt3} gives the distribution of the $p^2_T$ for the third leading jet for  both the background and the signal. Though this cut is not used in the results, it presents a possibility for an additional variable to be used in a multivariate analysis using Boosted Decision Tree or BDT \cite{Speckmayer:2010zz}.
	
\end{itemize}

\subsection{Results and Discussions}

Table \ref{results} gives the background and signal efficiencies at each level of the cuts imposed. Note that the $tWb$ final state corresponds to a $RS$ like 
scenario with singlet VLQ $T$. Though the efficiencies are quoted only for a particular final state for simplicity, the robustness of the analyses ensures that one 
can expect similar efficiencies for the other final state \textit{viz.} $tth,ttz$. Thus in a realistic scenario all possible decay modes of the VLQ are taken into 
account without any particular bias. 

Since the final state is associated with three hard parton which will eventually form jets, we require the presence of at least three jets in the event. Though it 
may seem that the presence of single isolated lepton may make the analysis less generic, it must be noted that the final state is always associated with the 
presence of a $t$: either from the VLQ or from the vertex of the decay of the coloured boson. As a result the lepton in this case can be attributed to the top. 
This requirement is particularly necessary to keep the large QCD background in check. The relaxing of this criteria would warrant the investigation of the 
substructure of these jets \cite{Chakraborty:2017mbz,Aguilar-Saavedra:2017rzt} to distinguish the signal from QCD and will not be considered here.

The large $p_T$ cut ($p_T>800$ GeV)is imposed to ensure that the final state jets have originated from a massive resonance. This is further complemented by 
the requirement of a minimum on the invariant mass of the three leading jets ($m_{j_0j_1j_2}>2000$ GeV). At this point it is important to note that these cuts 
are also very useful in keeping the `SUSY backgrounds' in check. Note that SUSY particles are typically produced in pairs and for final states from SUSY 
topologies to pass these basic, it requires the pair production of very massive SUSY particles. Subsequently, this results in a a significant drop in their production 
cross-section and much beyond the range of luminosities discussed here. 

Irrespective of the final state, both signal topologies exhibit similar efficiencies thus validating our analysis. The efficiencies for BP2 is likely to be slightly higher as 
the new physics states are more massive thereby resulting in more events passing the $p_T$ criteria. We now discuss the luminosity reach corresponding to the 
efficiencies quoted in Table \ref{results}. For brevity we discuss the numbers corresponding to the $ttg$ final state.

For BP1, the typical production cross-section for the coloured boson is 100 $fb^{-1}$.
 Further the B.R of the coloured boson ($C\rightarrow tT$) depends on the model. \cite{Chala:2014mma} considered different possibilities when the coloured 
 boson is a heavy gluon. Assuming a 60$\%$ branching fraction these channels can be probed at a $S/\sqrt{B}\sim 5\sigma$ sensitivity with 800 $fb^{-1}$ of 
 data. Extending  the luminosity to 3000 $fb^{-1}$ even models with $\sim 30\%$ branching ratios can be probed at a similar sensitivity.
 
 For BP2 the associated production cross-sections are much lower with typical values $\sim 30$ fb \cite{Iyer:2016yjb}. As a result, with $60\%$ B.R. of the coloured state into the $tT$ pair, $2400 fb^{-1}$ of data is required for $S/\sqrt{B}\sim 5\sigma$ sensitivity. In this case B.R. as low as $54\%$ can be probed with $3000 fb^{-1}$ of data.  
 
 We also comment on the case where the mass of VLQ for BP2 is increased to 3 TeV. In this case the efficiencies are similar to that for BP2 in Table \ref{results}. B.R. in this case are however likely to be smaller owing to the smaller mass difference:$\Delta m=m_C-m_T$.
  Smaller B.R. of upto $33\%$ can be probed with 3000 $fb^{-1}$ at $3\sigma$ sensitivity thereby making a strong case for future colliders.  \\
 
 \textbf{Mass of the vector like quark:} While the simple cuts in Table \ref{results} help us in identifying the specific topology under considerations, it is also 
 worthwhile to get a rough estimate of the mass of the VLQ. Since we are looking at complete jet final state, the identity of the partons from the VLQ is not 
 straightforward and hence one must consider different combinations of the invariant mass distributions. Right panel of Fig. \ref{invmass} gives the invariant mass of 
 the two leading jets $m_{j_0j_1}$ and for both the benchmark points the distribution terminates at roughly the mass of the VLQ. While this is useful, it can only 
 be used as a guideline. In practice one must consider other combinations to help in uniquely determining the mass.
\begin{table}[h]
	\begin{center}

		\begin{tabular}{|ccccccc|}
			
			\multicolumn{5}{c}{1 lepton + $n_J >=3$ } \\
			\hline
			&\multicolumn{2}{|c|}{Background Samples }&\multicolumn{2}{|c|}{Signal(BP1) }&\multicolumn{2}{|c|}{Signal(BP2) } \\
			
			\multicolumn{1}{ |c|  }{Cuts}&SM (1.74 pb)&$C\rightarrow t\bar t$  &\multicolumn{1}{ |c  }{$ttg$}&\multicolumn{1}{ c  }{$tWb$}&\multicolumn{1}{ |c  }{$ttg$}&\multicolumn{1}{ c|  }{$tWb$}\\
			\cline{1-7}
			\multicolumn{1}{ |c|  }{-}&0.18& 0.053 &\multicolumn{1}{ |c  }{0.16}&\multicolumn{1}{ c|  }{0.23}&\multicolumn{1}{ |c  }{$0.15$}&\multicolumn{1}{ c|  }{$0.23$}\\
			\multicolumn{1}{ |c|  }{$p^{(0)}_T>800$ GeV}&0.02&0.007  &\multicolumn{1}{ |c  }{0.027}&\multicolumn{1}{ c|  }{0.026}&\multicolumn{1}{ |c  }{$0.033$}&\multicolumn{1}{ c|  }{$0.029$}\\
			\multicolumn{1}{ |c|  }{$m_{j_0j_1j_2}>2000$ GeV}&0.013&0.0027  &\multicolumn{1}{ |c  }{0.014}&\multicolumn{1}{ c|  }{0.013}&\multicolumn{1}{ |c  }{$0.027$}&\multicolumn{1}{ c|  }{$0.017$}\\

%

			\hline
		\end{tabular}
	\end{center}
	\caption{Signal and Background efficiencies for different level of cuts employed}
	\label{results}
\end{table}

\section{Conclusions}
Coloured bosons are present in typical extensions of the SM and in particular in fundamental composite models and extra dimensional models. 
These models also typically contain VLQ states. We have investigated the possibility of producing VLQs in association with those coloured states using a 
generic effective Lagrangian framework taking into account the possible couplings based on the quantum numbers of those particles. In this respect the
analysis performed here can be therefore used for various different models. For the collider analysis at the LHC we have shown that a common strategy can 
be used for these resonances. Furthermore we focus our study to final states which are not specific and that can be broadly classified as jets, without 
searching for specific detailed signatures (for example containing leptons or reconstructing gauge boson masses). While this approach is typically less 
constraining for searches and establishing bounds, it is however quite robust and generic, allowing in this simplified study to used previous analyses and to apply 
the results to a wide class of models. The main outcome of the present study is that bounds on VLQs can be improved beyond the reach of standard searches 
if the production in association with heavy coloured bosons is considered. This reach can be extended to $\sim 3$ TeV in cases with heavy coloured state $m_c\sim3.5-3.7$ TeV  having a moderate to large $B.R.(C\rightarrow tT)$.
The hypothesis of extra coloured stated is motivated by their presence in a large 
number of models and is by no mean an exception. The presence of a colour octet is considered in the some of the existing searches, for 
example in the form of a heavy gluon, but more generally other quantum numbers are possible for these states (for example the sextet) and their phenomenology 
and implications are not fully explored. Their study in conjunction with VLQs was therefore one of the main motivations for the present study. Finally in the case 
that such events are seen at the LHC, we have briefly address the possibility of discriminating such a scenario from other models and in particular 
supersymmetric models which are characterised by the absence of VLQs. However the decay of coloured objects like stops or gluinos may produce in principle 
VLQ-like signals, either in terms of jet multiplicity or the number of isolated leptons. We have shown that jet multiplicity is different in those cases and also that 
other discriminating criteria can be used, such as the study of missing energy and the edges of distributions. The analysis can be further improved by considering using multivariate techniques thus providing better perspectives on the existance of these states.

\appendix
\section{Details of the simulation}
\label{app:A}
In this section we outline the details of the simulation and jet-clustering algorithm employed in our analysis.
\begin{itemize}
	\item The events are generated using {\tt{MADGRAPH}} \cite{Alwall:2014hca} at 13 TeV CM energy using PDF NNLO1. The model file for the processes is generated using {\tt{FEYNRULES}} \cite{Christensen:2008py}. The events are then showered and hadronized using {\tt{PYTHIA 8}} \cite{Sjostrand:2007gs} and then passed to the {\tt{DELPHES 3.3.2}} \cite{deFavereau:2013fsa} detector simulator using the CMS card.
	
	\item The jets are clustered with the particle-flow objects using {\tt{FASTJET}} \cite{Cacciari:2011ma} with the $anti$-$kt$ \cite{Cacciari:2008gp} with the following parameters: $R=0.8$ and $p^{min}_T=100$ GeV.
	
	\item The extracted jets used for the analysis are not subject to the Unique object finder module to distingush it from the isolated leptons. This implies that the jets now include even those initiated by the leptons. 
	
	\item Requirement of a single isolated lepton is only imposed at the event selection criteria to reject the QCD background. The standard parameters for lepton isolation are chosen: isolation radius of $R=0.4$ and the $p_T$ fraction of the activity around lepton canditate to be less than $0.1$ of the $p_T$ of the lepton.
\end{itemize}

\paragraph{Acknowledgements}
We would like to acknowledge many useful discussions with Ushoshi Maitra and her collaboration in the initial stages of the project.
AD is partially supported by the ``Institut Universitaire de France''; the Labex-LIO (Lyon Institute of 
Origins) under grant ANR-10-LABX-66 and FRAMA (FR3127, F\'ed\'eration de Recherche ``Andr\'e Marie Amp\`ere"). 
A.I. was supported in part by MIUR under Project No. 2015P5SBHT and by the INFN research initiative ENP. 
A.I. would like to acknowledge the hospitality at IPN Lyon where significant parts of the project were discussed. We would also like to acknowledge the organisers of From Strings to LHC IV where parts of the project were discussed. AD and AI would also like to thank organisers of Les Houches Workshop series 2017 where part of the discussions were held. We would like to acknowledge the computational facilities at Department of Theoretical Physics, TIFR. 

We  would like to warmly acknowledge the support of the CNRS LIA (Laboratoire International Associ\'e) THEP (Theoretical High Energy Physics) and 
the INFRE-HEPNET (IndoFrench Network on High Energy Physics) of CEFIPRA/IFCPAR (Indo-French Centre for the Promotion of Advanced Research).

\bibliography{biblio}

\providecommand{\href}[2]{#2}\begingroup\raggedright\begin{thebibliography}{10}

\bibitem{Lewis:2011zb}
R.~Lewis, C.~Pica, and F.~Sannino, {\it {Light Asymmetric Dark Matter on the
  Lattice: SU(2) Technicolor with Two Fundamental Flavors}},  {\em Phys. Rev.}
  {\bf D85} (2012) 014504, [\href{http://xxx.lanl.gov/abs/1109.3513}{{\tt
  arXiv:1109.3513}}].

\bibitem{Cacciapaglia:2015eqa}
G.~Cacciapaglia, H.~Cai, A.~Deandrea, T.~Flacke, S.~J. Lee, and A.~Parolini,
  {\it {Composite scalars at the LHC: the Higgs, the Sextet and the Octet}},
  {\em JHEP} {\bf 11} (2015) 201,
  [\href{http://xxx.lanl.gov/abs/1507.02283}{{\tt arXiv:1507.02283}}].

\bibitem{Randall:1999ee}
L.~Randall and R.~Sundrum, {\it {A Large mass hierarchy from a small extra
  dimension}},  {\em Phys. Rev. Lett.} {\bf 83} (1999) 3370--3373,
  [\href{http://xxx.lanl.gov/abs/hep-ph/9905221}{{\tt hep-ph/9905221}}].

\bibitem{Gherghetta:2000qt}
T.~Gherghetta and A.~Pomarol, {\it {Bulk fields and supersymmetry in a slice of
  AdS}},  {\em Nucl. Phys.} {\bf B586} (2000) 141--162,
  [\href{http://xxx.lanl.gov/abs/hep-ph/0003129}{{\tt hep-ph/0003129}}].

\bibitem{Agashe:2006hk}
K.~Agashe, A.~Belyaev, T.~Krupovnickas, G.~Perez, and J.~Virzi, {\it {LHC
  Signals from Warped Extra Dimensions}},  {\em Phys. Rev.} {\bf D77} (2008)
  015003, [\href{http://xxx.lanl.gov/abs/hep-ph/0612015}{{\tt
  hep-ph/0612015}}].

\bibitem{Lillie:2007yh}
B.~Lillie, L.~Randall, and L.-T. Wang, {\it {The Bulk RS KK-gluon at the LHC}},
   {\em JHEP} {\bf 09} (2007) 074,
  [\href{http://xxx.lanl.gov/abs/hep-ph/0701166}{{\tt hep-ph/0701166}}].

\bibitem{Guchait:2007jd}
M.~Guchait, F.~Mahmoudi, and K.~Sridhar, {\it {Associated production of a
  Kaluza-Klein excitation of a gluon with a t anti-t pair at the LHC}},  {\em
  Phys. Lett.} {\bf B666} (2008) 347--351,
  [\href{http://xxx.lanl.gov/abs/0710.2234}{{\tt arXiv:0710.2234}}].

\bibitem{Iyer:2016yjb}
A.~M. Iyer, F.~Mahmoudi, N.~Manglani, and K.~Sridhar, {\it {Kaluza–Klein
  gluon + jets associated production at the Large Hadron Collider}},  {\em
  Phys. Lett.} {\bf B759} (2016) 342--348,
  [\href{http://xxx.lanl.gov/abs/1601.02033}{{\tt arXiv:1601.02033}}].

\bibitem{Han:2003wu}
T.~Han, H.~E. Logan, B.~McElrath, and L.-T. Wang, {\it {Phenomenology of the
  little Higgs model}},  {\em Phys. Rev.} {\bf D67} (2003) 095004,
  [\href{http://xxx.lanl.gov/abs/hep-ph/0301040}{{\tt hep-ph/0301040}}].

\bibitem{Contino:2006qr}
R.~Contino, L.~Da~Rold, and A.~Pomarol, {\it {Light custodians in natural
  composite Higgs models}},  {\em Phys. Rev.} {\bf D75} (2007) 055014,
  [\href{http://xxx.lanl.gov/abs/hep-ph/0612048}{{\tt hep-ph/0612048}}].

\bibitem{Carena:2006jx}
M.~Carena, J.~Hubisz, M.~Perelstein, and P.~Verdier, {\it {Collider signature
  of T-quarks}},  {\em Phys. Rev.} {\bf D75} (2007) 091701,
  [\href{http://xxx.lanl.gov/abs/hep-ph/0610156}{{\tt hep-ph/0610156}}].

\bibitem{Matsumoto:2008fq}
S.~Matsumoto, T.~Moroi, and K.~Tobe, {\it {Testing the Littlest Higgs Model
  with T-parity at the Large Hadron Collider}},  {\em Phys. Rev.} {\bf D78}
  (2008) 055018, [\href{http://xxx.lanl.gov/abs/0806.3837}{{\tt
  arXiv:0806.3837}}].

\bibitem{Anastasiou:2009rv}
C.~Anastasiou, E.~Furlan, and J.~Santiago, {\it {Realistic Composite Higgs
  Models}},  {\em Phys. Rev.} {\bf D79} (2009) 075003,
  [\href{http://xxx.lanl.gov/abs/0901.2117}{{\tt arXiv:0901.2117}}].

\bibitem{Kribs:2010ii}
G.~D. Kribs, A.~Martin, and T.~S. Roy, {\it {Higgs boson discovery through
  top-partners decays using jet substructure}},  {\em Phys. Rev.} {\bf D84}
  (2011) 095024, [\href{http://xxx.lanl.gov/abs/1012.2866}{{\tt
  arXiv:1012.2866}}].

\bibitem{Gopalakrishna:2011ef}
S.~Gopalakrishna, T.~Mandal, S.~Mitra, and R.~Tibrewala, {\it {LHC Signatures
  of a Vector-like b'}},  {\em Phys. Rev.} {\bf D84} (2011) 055001,
  [\href{http://xxx.lanl.gov/abs/1107.4306}{{\tt arXiv:1107.4306}}].

\bibitem{DeSimone:2012fs}
A.~De~Simone, O.~Matsedonskyi, R.~Rattazzi, and A.~Wulzer, {\it {A First Top
  Partner Hunter's Guide}},  {\em JHEP} {\bf 04} (2013) 004,
  [\href{http://xxx.lanl.gov/abs/1211.5663}{{\tt arXiv:1211.5663}}].

\bibitem{Vignaroli:2012sf}
N.~Vignaroli, {\it {Discovering the composite Higgs through the decay of a
  heavy fermion}},  {\em JHEP} {\bf 07} (2012) 158,
  [\href{http://xxx.lanl.gov/abs/1204.0468}{{\tt arXiv:1204.0468}}].

\bibitem{Gopalakrishna:2013hua}
S.~Gopalakrishna, T.~Mandal, S.~Mitra, and G.~Moreau, {\it {LHC Signatures of
  Warped-space Vectorlike Quarks}},  {\em JHEP} {\bf 08} (2014) 079,
  [\href{http://xxx.lanl.gov/abs/1306.2656}{{\tt arXiv:1306.2656}}].

\bibitem{Buchkremer:2013bha}
M.~Buchkremer, G.~Cacciapaglia, A.~Deandrea, and L.~Panizzi, {\it {Model
  Independent Framework for Searches of Top Partners}},  {\em Nucl. Phys.} {\bf
  B876} (2013) 376--417, [\href{http://xxx.lanl.gov/abs/1305.4172}{{\tt
  arXiv:1305.4172}}].

\bibitem{Banfi:2013yoa}
A.~Banfi, A.~Martin, and V.~Sanz, {\it {Probing top-partners in Higgs+jets}},
  {\em JHEP} {\bf 08} (2014) 053,
  [\href{http://xxx.lanl.gov/abs/1308.4771}{{\tt arXiv:1308.4771}}].

\bibitem{Li:2013xba}
J.~Li, D.~Liu, and J.~Shu, {\it {Towards the fate of natural composite Higgs
  model through single $t^\prime$ search at the 8 TeV LHC}},  {\em JHEP} {\bf
  11} (2013) 047, [\href{http://xxx.lanl.gov/abs/1306.5841}{{\tt
  arXiv:1306.5841}}].

\bibitem{Gripaios:2014pqa}
B.~Gripaios, T.~Müller, M.~A. Parker, and D.~Sutherland, {\it {Search
  Strategies for Top Partners in Composite Higgs models}},  {\em JHEP} {\bf 08}
  (2014) 171, [\href{http://xxx.lanl.gov/abs/1406.5957}{{\tt
  arXiv:1406.5957}}].

\bibitem{Chala:2014mma}
M.~Chala, J.~Juknevich, G.~Perez, and J.~Santiago, {\it {The Elusive Gluon}},
  {\em JHEP} {\bf 01} (2015) 092,
  [\href{http://xxx.lanl.gov/abs/1411.1771}{{\tt arXiv:1411.1771}}].

\bibitem{Endo:2014bsa}
M.~Endo, K.~Hamaguchi, K.~Ishikawa, and M.~Stoll, {\it {Reconstruction of
  Vector-like Top Partner from Fully Hadronic Final States}},  {\em Phys. Rev.}
  {\bf D90} (2014), no.~5 055027,
  [\href{http://xxx.lanl.gov/abs/1405.2677}{{\tt arXiv:1405.2677}}].

\bibitem{Dolan:2016eki}
M.~J. Dolan, J.~L. Hewett, M.~Krämer, and T.~G. Rizzo, {\it {Simplified Models
  for Higgs Physics: Singlet Scalar and Vector-like Quark Phenomenology}},
  {\em JHEP} {\bf 07} (2016) 039,
  [\href{http://xxx.lanl.gov/abs/1601.07208}{{\tt arXiv:1601.07208}}].

\bibitem{Deandrea:2014raa}
A.~Deandrea and N.~Deutschmann, {\it {Multi-tops at the LHC}},  {\em JHEP} {\bf
  08} (2014) 134, [\href{http://xxx.lanl.gov/abs/1405.6119}{{\tt
  arXiv:1405.6119}}].

\bibitem{Sirunyan:2017bfa}
{\bf CMS} Collaboration, A.~M. Sirunyan et~al., {\it {Search for a heavy
  resonance decaying to a top quark and a vector-like top quark at $\sqrt{s}$ =
  13 TeV}},  \href{http://xxx.lanl.gov/abs/1703.06352}{{\tt arXiv:1703.06352}}.

\bibitem{Iyer:2016qzd}
A.~M. Iyer and U.~Maitra, {\it {Dissecting new physics models through kinematic
  edges}},  {\em Phys. Rev.} {\bf D95} (2017), no.~3 035039,
  [\href{http://xxx.lanl.gov/abs/1609.06502}{{\tt arXiv:1609.06502}}].

\bibitem{Araque:2015cna}
J.~P. Araque, N.~F. Castro, and J.~Santiago, {\it {Interpretation of
  Vector-like Quark Searches: Heavy Gluons in Composite Higgs Models}},  {\em
  JHEP} {\bf 11} (2015) 120, [\href{http://xxx.lanl.gov/abs/1507.05628}{{\tt
  arXiv:1507.05628}}].

\bibitem{Agashe:2004rs}
K.~Agashe, R.~Contino, and A.~Pomarol, {\it {The Minimal composite Higgs
  model}},  {\em Nucl. Phys.} {\bf B719} (2005) 165--187,
  [\href{http://xxx.lanl.gov/abs/hep-ph/0412089}{{\tt hep-ph/0412089}}].

\bibitem{Chala:2017xgc}
M.~Chala, {\it {Direct bounds on heavy toplike quarks with standard and exotic
  decays}},  {\em Phys. Rev.} {\bf D96} (2017), no.~1 015028,
  [\href{http://xxx.lanl.gov/abs/1705.03013}{{\tt arXiv:1705.03013}}].

\bibitem{Aad:2016shx}
{\bf ATLAS} Collaboration, G.~Aad et~al., {\it {Search for single production of
  a vector-like quark via a heavy gluon in the $4b$ final state with the ATLAS
  detector in $pp$ collisions at $\sqrt{s} = 8$ TeV}},  {\em Phys. Lett.} {\bf
  B758} (2016) 249--268, [\href{http://xxx.lanl.gov/abs/1602.06034}{{\tt
  arXiv:1602.06034}}].

\bibitem{Speckmayer:2010zz}
P.~Speckmayer, A.~Hocker, J.~Stelzer, and H.~Voss, {\it {The toolkit for
  multivariate data analysis, TMVA 4}},  {\em J. Phys. Conf. Ser.} {\bf 219}
  (2010) 032057.

\bibitem{Chakraborty:2017mbz}
A.~Chakraborty, A.~M. Iyer, and T.~S. Roy, {\it {A Universal Framework for
  Finding Anomalous Objects at the LHC}},
  \href{http://xxx.lanl.gov/abs/1707.07084}{{\tt arXiv:1707.07084}}.

\bibitem{Aguilar-Saavedra:2017rzt}
J.~A. Aguilar-Saavedra, J.~H. Collins, and R.~K. Mishra, {\it {A generic
  anti-QCD jet tagger}},  \href{http://xxx.lanl.gov/abs/1709.01087}{{\tt
  arXiv:1709.01087}}.

\bibitem{Alwall:2014hca}
J.~Alwall, R.~Frederix, S.~Frixione, V.~Hirschi, F.~Maltoni, O.~Mattelaer,
  H.~S. Shao, T.~Stelzer, P.~Torrielli, and M.~Zaro, {\it {The automated
  computation of tree-level and next-to-leading order differential cross
  sections, and their matching to parton shower simulations}},  {\em JHEP} {\bf
  07} (2014) 079, [\href{http://xxx.lanl.gov/abs/1405.0301}{{\tt
  arXiv:1405.0301}}].

\bibitem{Christensen:2008py}
N.~D. Christensen and C.~Duhr, {\it {FeynRules - Feynman rules made easy}},
  {\em Comput.Phys.Commun.} {\bf 180} (2009) 1614--1641,
  [\href{http://xxx.lanl.gov/abs/0806.4194}{{\tt arXiv:0806.4194}}].

\bibitem{Sjostrand:2007gs}
T.~Sjostrand, S.~Mrenna, and P.~Z. Skands, {\it {A Brief Introduction to PYTHIA
  8.1}},  {\em Comput. Phys. Commun.} {\bf 178} (2008) 852--867,
  [\href{http://xxx.lanl.gov/abs/0710.3820}{{\tt arXiv:0710.3820}}].

\bibitem{deFavereau:2013fsa}
{\bf DELPHES 3} Collaboration, J.~de~Favereau, C.~Delaere, P.~Demin,
  A.~Giammanco, V.~Lemaître, A.~Mertens, and M.~Selvaggi, {\it {DELPHES 3, A
  modular framework for fast simulation of a generic collider experiment}},
  {\em JHEP} {\bf 02} (2014) 057,
  [\href{http://xxx.lanl.gov/abs/1307.6346}{{\tt arXiv:1307.6346}}].

\bibitem{Cacciari:2011ma}
M.~Cacciari, G.~P. Salam, and G.~Soyez, {\it {FastJet User Manual}},  {\em Eur.
  Phys. J.} {\bf C72} (2012) 1896,
  [\href{http://xxx.lanl.gov/abs/1111.6097}{{\tt arXiv:1111.6097}}].

\bibitem{Cacciari:2008gp}
M.~Cacciari, G.~P. Salam, and G.~Soyez, {\it {The Anti-k(t) jet clustering
  algorithm}},  {\em JHEP} {\bf 04} (2008) 063,
  [\href{http://xxx.lanl.gov/abs/0802.1189}{{\tt arXiv:0802.1189}}].

\end{thebibliography}\endgroup
\bibliographystyle{JHEP}

\end{document}